# Magnetic Resonance Image Processing Transformer for General Accelerated Image Reconstruction

*Guoyao Shen†, Mengyu Li†, Stephan Anderson, Chad W. Farris, and Xin Zhang\**

*†* Equal contribution

G. Shen, M. Li, X. Zhang

Department of Mechanical Engineering, Boston University, Boston, MA, 02215 USA

E-mail: xinz@bu.edu

G. Shen, M. Li, S. Anderson, X. Zhang

Photonics Center, Boston University, Boston, MA, 02215 USA

S. Anderson, C. W. Farris

Chobanian & Avedisian School of Medicine, Boston University Medical Campus, Boston, MA, 02118 USA

X. Zhang

Department of Electrical & Computer Engineering, Boston, MA, 02215 USA

Department of Biomedical Engineering, Boston, MA, 02215 USA

Division of Materials Science & Engineering, Boston, MA, 02215 USA

Rafik B. Hariri Institute for Computing and Computational Science & Engineering, Boston, MA, 02215 USA

Abstract:

Recent advancements in deep learning have enabled the development of generalizable models that achieve state-of-the-art performance across various imaging tasks. Vision Transformer (ViT)-based architectures, in particular, have demonstrated strong feature extraction capabilities when pre-trained on large-scale datasets. In this work, we introduce the Magnetic Resonance Image Processing Transformer (MR-IPT), a ViT-based framework designed to enhance the generalizability and robustness of accelerated MRI reconstruction. Unlike conventional deep learning models that require separate training for different acceleration factors, MR-IPT is pre-trained on a large-scale dataset encompassing multiple undersampling patterns and acceleration settings, enabling a unified reconstruction framework. By leveraging a shared transformer backbone, MR-IPT effectively learns universal feature representations, allowing it to generalize across diverse reconstruction tasks. Extensive experiments demonstrate that MR-IPT outperforms both CNN-based and existing transformer-based methods, achieving superior reconstruction quality across varying acceleration factors and sampling masks. Moreover, MR-IPT exhibits strong robustness, maintaining high performance even under unseen acquisition setups, highlighting its potential as a scalable and efficient solution for accelerated MRI. Our findings suggest that transformer-based general models can significantly advance MRI reconstruction, offering improved adaptability and stability compared to traditional deep learning approaches.

## 1. Introduction

Magnetic Resonance Imaging (MRI) is a widely used diagnostic and research tool in clinical settings, offering high-resolution imaging and diverse contrast mechanisms to visualize various structural and functional characteristics of the underlying anatomy. However, one of the significant limitations of MRI is its relatively long acquisition time, which can reduce patient throughput, increase costs, and lead to delays in diagnosis. Long scan times can also contribute to patient discomfort and motion-related artifacts, which may degrade image quality. To address this challenge, various acceleration techniques have been developed, with compressed sensing (CS)[1-3] being one of the most widely adopted methods. CS accelerates MRI acquisition by undersampling k-space data, relying on sparsity constraints and iterative reconstruction algorithms to recover high-quality images from limited measurements. While effective, CS-based approaches often introduce reconstruction errors, including aliasing artifacts and loss of fine structural details, particularly in high-acceleration settings.

Recent advances in deep learning have provided powerful alternatives for accelerated MRI reconstruction.[4-10] Deep learning-based methods leverage large datasets to learn complex mappings between undersampled and fully sampled k-space or image domain representations. Many state-of-the-art deep learning models for MRI reconstruction are based on convolutional neural networks (CNNs). These CNN-based architectures have demonstrated remarkable improvements over traditional CS methods, yielding higher-quality reconstructions with fewer artifacts and faster inference times. However, CNNs have inherent limitations due to their local receptive fields and translation-invariant convolutional operations.[11, 12] These characteristics can restrict their ability to capture long-range dependencies and global contextual information, leading to suboptimal reconstruction performance, especially in cases where high-frequency details are critical.[13]

The emergence of transformer architectures has introduced new possibilities for image processing and computer vision tasks. Originally developed for natural language processing (NLP), transformers[14] and their variants have been widely applied to tasks such as text classification, machine translation, and question-answering.[15-20] A key advantage of transformer-based models is their ability to capture long-range dependencies and contextual information via self-attention mechanisms.[21, 22] Inspired by this, Vision Transformer (ViT)[23] adapted transformers for image-related tasks by treating input images as sequences of non-overlapping patches, similar to words in NLP. Unlike CNNs, which gradually expands the receptive field hierarchically, even a shallow ViT model can effectively model global contextual relationships, making it highly competitive across various vision applications.[24-26]

Further developments in transformer-based architectures have led to innovations such as masked autoencoders (MAE),[27] which utilize a masked token prediction strategy and encoder-decoder structure during pretraining to enhance representation learning. These techniques have demonstrated strong potential in image reconstruction tasks, where learning robust feature representations is crucial. Additionally, models like the Segment

Anything Model (SAM)[28, 29] have demonstrated the adaptability of ViT-based structures as backbones for multimodal learning, further expanding their applicability to medical imaging.

Despite these advancements, applying transformer-based models to accelerated MRI reconstruction remains an area of active research. Several studies have explored ViT-based architectures for MRI reconstruction, showing that transformer models can benefit from pretraining on large datasets and outperform CNN-based approaches in certain settings.[30, 31] However, most of these efforts focus on task-specific models rather than developing a generalized framework for MRI reconstruction.[32] Existing ViT-based models are often designed for specific undersampling patterns and acceleration factors, limiting their adaptability across different acquisition setups. A more generalizable approach is needed to fully leverage the capabilities of transformers in MRI reconstruction.

Image Processing Transformer (IPT)[33] has emerged as a promising framework for achieving generalizability in low-level imaging tasks. IPT introduces a multi-task learning paradigm by incorporating multiple input-output configurations within a single framework. By integrating multiple heads and tails for different image processing tasks, the shared transformer body learns to extract universal feature representations, improving model adaptability across diverse imaging scenarios. This design has demonstrated success in generalizing across multiple low-level image processing tasks such as denoising, deraining, and super-resolution.

Motivated by these developments, we introduce the Magnetic Resonance Image Processing Transformer (MR-IPT), a novel framework designed to enhance the generalizability of ViT-based models for accelerated MRI reconstruction. MR-IPT extends the IPT paradigm by interpreting different undersampling reconstruction setups as distinct tasks, allowing the core ViT backbone to focus on learning robust feature representations. We pre-train MR-IPT on a large-scale medical imaging dataset to maximize its feature extraction capabilities. Subsequently, we evaluate its performance on multiple downstream MRI reconstruction tasks, incorporating various acceleration factors and sampling masks to assess its adaptability. Our experimental results demonstrate that MR-IPT outperforms both CNN- and ViT-based models across a range of MRI reconstruction scenarios. Notably, MR-IPT exhibits strong generalization capabilities, effectively handling unseen sampling patterns and acceleration rates. Additionally, we conduct model stability assessments, showing that MR-IPT maintains high reconstruction quality even when trained with limited downstream data. These findings highlight the potential of MR-IPT as a robust and scalable solution for accelerated MRI reconstruction.

## 2. Results

### 2.1. MR-IPT Framework

The MR-IPT framework consists of five core components: heads, tails, a prompt encoder, a shared encoder, and a shared decoder (**Figure 1a**). The shared encoder utilizes shifted-window multi-head self-attention (W-MSA)[34] to efficiently capture global context across multiple layers. Inspired by MAE and SAM, we implemented a lightweight decoder incorporating prompt self-attention and two-way cross-attention, facilitating effective feature refinement and reconstruction. This lightweight design allows for a deeper encoder architecture without significantly increasing model size and computational costs, thereby enhancing the model's representational capacity. The heads extract features from undersampled images, transforming them into patch tokens. The prompt encoder generates prompt tokens based on acceleration labels, which, together with image patch tokens, are processed by the shared encoder-decoder to recover missing information. The tails then reconstruct fully sampled images from the learned reconstruction tokens (**Figure 1b**). To ensure broad generalizability in accelerated MRI reconstruction, we train MR-IPT on images corrupted using a diverse range of sampling masks and acceleration ratios, as illustrated in **Figure 1c**.

Unlike the original IPT, which assigns a dedicated head-tail pair to each specific reconstruction task, we implement three MR-IPT variants across diverse undersampling patterns: (1) MR-IPT-type: Heads and tails are aggregated based on acceleration ratios, where each head-tail pair specializes in reconstructing images from different sampling masks; (2) MR-IPT-level: Heads and tails are aggregated based on sampling masks, allowing each head-tail pair to focus on reconstruction across different acceleration ratios; (3) MR-IPT-split: Each unique combination of sampling mask and acceleration ratio is assigned a dedicated head-tail pair.

To fully leverage the generalization potential of MR-IPT, we trained our model on RadImageNet,[35] a large-scale medical imaging dataset. Training images were corrupted at five levels, with sampling ratios ranging from two to ten, incorporating both 1D and 2D sampling masks to enhance robustness. For evaluation, we conducted downstream MRI reconstruction experiments on the fastMRI dataset,[36] assessing MR-IPT's performance across typical reconstruction scenarios, unseen sampling ratios and masks, zero-shot generalization capabilities, and model stability under limited data conditions.

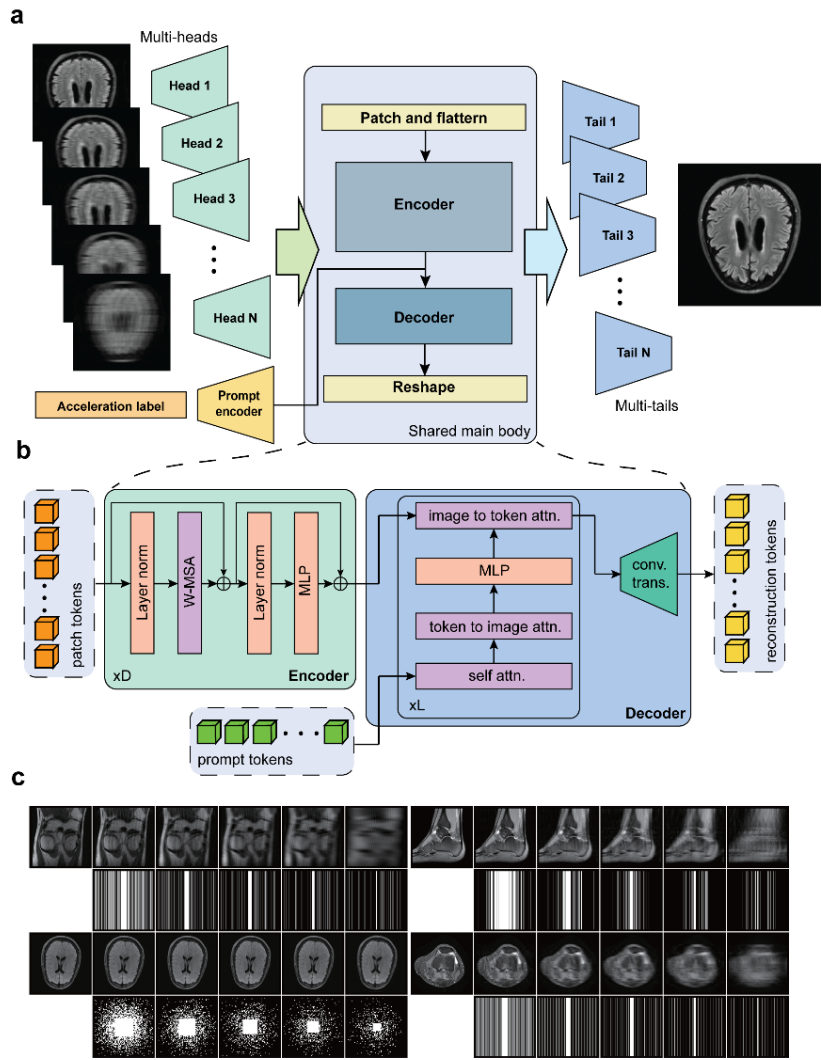

**Figure 1.** The overall architecture of the Magnetic Resonance Image Processing Transformer (MR-IPT) framework. a) MR-IPT consists of heads, tails, a prompt encoder, a shared encoder, and a shared decoder. b) A lightweight decoder incorporating prompt self-attention and two-way cross-attention enhances feature refinement and reconstruction. This design enables a deeper encoder architecture without significantly increasing model size or computational cost, improving the model's representational capacity. The heads extract features from undersampled images and transform them into patch tokens. The prompt encoder generates prompt tokens based on acceleration labels, which, together with image patch tokens, are processed by the shared encoder-decoder to recover missing information. The tails reconstruct fully sampled images from the learned reconstruction tokens. c) During pre-training, MR-IPT is trained on diverse acceleration ratios (2×, 4×, 6×, 8×, and 10×) and various sampling masks. The 1D sampling masks include Cartesian random, Cartesian equispaced, and 1D Gaussian, while the 2D sampling mask follows a 2D Gaussian distribution.

## 2.2. Accelerated MRI Reconstruction Performance

In this section, we evaluate MR-IPT on accelerated MRI reconstruction tasks using the fastMRI knee and brain datasets. Following the standard fastMRI benchmark, knee dataset reconstructions are performed with 4× and 8× Cartesian random undersampling, while brain dataset reconstructions are tested at 4× and 8× Cartesian equispaced undersampling.

We compare MR-IPT against multiple representative models, including UNet32,[37] UNet128, ViT-Base, and ViT-L.[23, 30] Additionally, to assess the impact of large-scale pretraining, we introduce UNet128-FT and ViT-L-FT, both of which are pretrained on RadImageNet—the same as MR-IPT—before being fine-tuned on the corresponding fastMRI datasets. The quantitative results are presented in **Table 1**, where MR-IPT consistently outperforms baseline models across different acceleration ratios. Among its three variants, MR-IPT-level exhibits slightly superior performance compared to MR-IPT-type and MR-IPT-split, suggesting that aggregating heads and tails by sampling mask provides an effective balance between task-specific optimization and generalizability.

For instance, in the 4× brain reconstruction task, MR-IPT-level achieves a PSNR/SSIM of 42.48/0.9831, significantly outperforming UNet128 (36.25/0.9648) and ViT-L (37.54/0.9558). Even when compared to pretrained and fine-tuned models (UNet128-FT: 37.27/0.9653, ViT-L-FT: 37.54/0.9558), MR-IPT still demonstrates a clear advantage. This highlights the effectiveness of MR-IPT's multi-head-tail structure and its unified shared encoder-decoder, which maximizes the benefits of large-scale pretraining by improving feature adaptability across different undersampling conditions.

**Figure 2** illustrates qualitative comparisons of reconstructed images, including error maps that represent absolute differences between reconstructions and ground truth images (intensified by a factor of three for better visualization). **Figure 3** provides a comparative analysis of the three MR-IPT variants (MR-IPT-type, MR-IPT-level, and MR-IPT-split) under 4× and 8× Cartesian random and Cartesian equispaced undersampling. Overall, MR-IPT produces cleaner error maps across all tested sampling ratios and masks, preserving finer anatomical structures with enhanced contrast fidelity. These results demonstrate MR-IPT's ability to achieve more accurate and perceptually superior MRI reconstructions, further validating its robustness and generalization capabilities.

**Table 1.** Reconstruction performance comparison on the fastMRI knee and brain datasets. We report peak signal-to-noise ratio (PSNR) and structural similarity index (SSIM) on the test sets, evaluating the quality of reconstructed images against fully sampled ground truth images.

|  | Knee | | | | Brain | | | |
|---|---|---|---|---|---|---|---|---|
|  | ACC=4X | | ACC=8X | | ACC=4X | | ACC=8X | |
|  | PSNR [dB] | SSIM | PSNR | SSIM | PSNR | SSIM | PSNR | SSIM |
| UNet32 | 31.86 | 0.8016 | 27.77 | 0.7311 | 35.66 | 0.9566 | 31.47 | 0.9231 |
| UNet128 | 32.12 | 0.8315 | 28.63 | 0.7462 | 36.25 | 0.9648 | 31.84 | 0.9243 |
| UNet128-FT | 32.55 | 0.8380 | 29.31 | 0.7566 | 37.27 | 0.9653 | 32.03 | 0.9248 |
| ViT-Base | 30.67 | 0.7797 | 27.88 | 0.7057 | 37.29 | 0.9547 | 33.01 | 0.9252 |
| ViT-L | 31.53 | 0.7942 | 28.67 | 0.7231 | 37.54 | 0.9558 | 33.04 | 0.9249 |
| ViT-L-FT | 32.09 | 0.8032 | 29.33 | 0.7375 | 37.63 | 0.9564 | 33.78 | 0.9318 |
| MR-IPT-*type* | 34.47 | 0.8671 | 31.38 | 0.7942 | 42.31 | 0.9827 | 35.34 | 0.9543 |
| MR-IPT-*level* | **34.52** | **0.8681** | **31.45** | **0.7952** | **42.48** | **0.9831** | **35.53** | **0.9557** |
| MR-IPT-*split* | 34.51 | 0.8678 | 31.44 | 0.7948 | 42.06 | 0.9827 | 35.34 | 0.9543 |

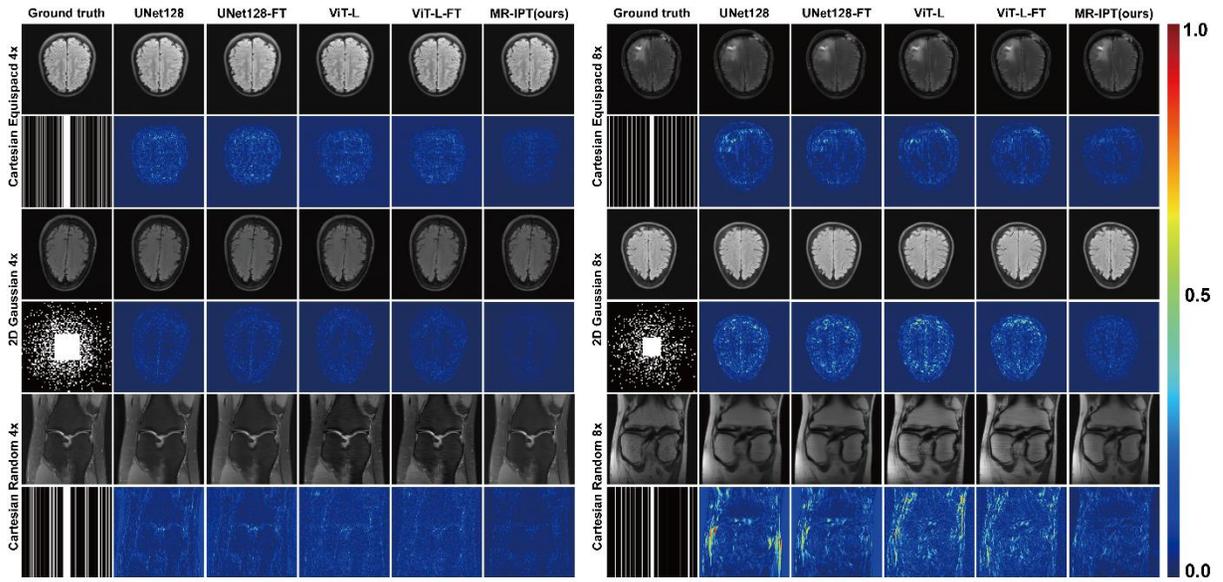

**Figure 2.** Reconstruction comparison across different models. Each row presents reconstructed images from various methods, highlighting differences in image quality and artifact suppression. The second row of each subplot shows the corresponding error maps (intensified by a factor of three for better visualization), which visualize absolute differences between the reconstructed images and the fully sampled ground truth.

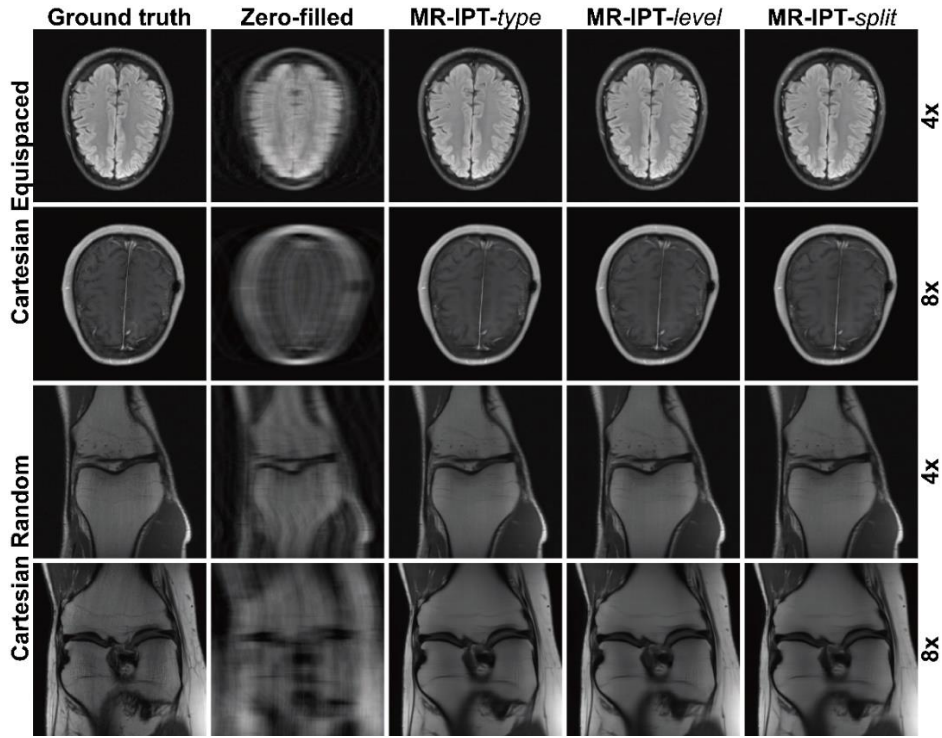

**Figure 3.** Comparison of MR-IPT variants across different undersampling patterns. We implement three MR-IPT variants across diverse undersampling patterns: (1) MR-IPT-type, where heads and tails are grouped based on acceleration ratios, with each head-tail pair specializing in different sampling masks; (2) MR-IPT-level, where heads and tails are grouped based on sampling masks, allowing each pair to generalize across different acceleration ratios; and (3) MR-IPT-split, where each unique combination of sampling mask and acceleration ratio is assigned a dedicated head-tail pair. The results demonstrate that all three variants achieve high-quality reconstructions, highlighting MR-IPT's flexibility in handling diverse undersampling patterns.

## 2.3 Performance on New Sampling Ratios

Given that downstream tasks may involve different undersampling ratios, it is crucial to assess MR-IPT's generalization to previously unseen acceleration factors. To keep a balance between the number of head-tail pairs and overall generalizability, we pre-train MR-IPT using five acceleration ratios ($2\times$, $4\times$, $6\times$, $8\times$, and $10\times$), covering a broad range of undersampling scenarios. To further evaluate its adaptability, we conduct downstream reconstruction on the brain dataset with unseen acceleration ratios of $5\times$ and $7\times$ during inference.

**Table 2** presents the quantitative results. UNet128 and ViT-L are trained directly on the indicated acceleration ratios, whereas UNet128-FT and ViT-L-FT follow the same pretraining-finetuning pipeline as MR-IPT for a fair comparison. Notably, both UNet128 and ViT-L benefit from pretraining, with ViT-L-FT showing greater improvements, achieving PSNR/SSIM of 36.29/0.9490 ($5\times$) and 34.60/0.9378 ($7\times$), compared to UNet128-FT at 35.52/0.9554 ($5\times$) and 32.96/0.9346 ($7\times$). MR-IPT-level consistently outperforms both models, achieving PSNR/SSIM of 39.92/0.9763 ($5\times$) and 36.69/0.9626 ($7\times$), highlighting its superior reconstruction capability.

**Figure 4** provides qualitative comparisons of reconstructed images. Interestingly, while ViT-L-FT and UNet128-FT exhibit differences in error map characteristics, UNet128-FT, despite having a higher maximum absolute error, produces cleaner backgrounds, particularly at higher acceleration ratios where overall image intensity is lower. MR-IPT consistently yields the cleanest error maps across all tested setups, including unseen sampling ratios, demonstrating its strong adaptability and generalization to novel acceleration factors.

**Table 2.** Reconstruction performance comparison on unseen sampling ratios (5× and 7×) for the brain dataset. UNet128-FT and ViT-L-FT follow the same pretraining-finetuning pipeline as MR-IPT to ensure a fair evaluation of generalization performance. Results demonstrate MR-IPT's superior adaptability to novel acceleration factors, outperforming both UNet128-FT and ViT-L-FT in PSNR and SSIM.

|  | Brain – Cartesian Equispaced | | | |
|---|---|---|---|---|
|  | ACC=5X | | ACC=7X | |
|  | PSNR | SSIM | PSNR | SSIM |
| UNet128 | 35.20 | 9.9554 | 32.66 | 0.9339 |
| UNet128-FT | 35.52 | 0.9554 | 32.96 | 0.9346 |
| ViT-L | 35.58 | 0.9447 | 33.60 | 0.9296 |
| ViT-L-FT | 36.29 | 0.9490 | 34.60 | 0.9378 |
| MR-IPT-*type* | 39.71 | 0.9757 | 36.56 | 0.9619 |
| MR-IPT-*level* | **39.92** | **0.9763** | **36.69** | **0.9626** |
| MR-IPT-*split* | 39.78 | 0.9758 | 36.49 | 0.9615 |

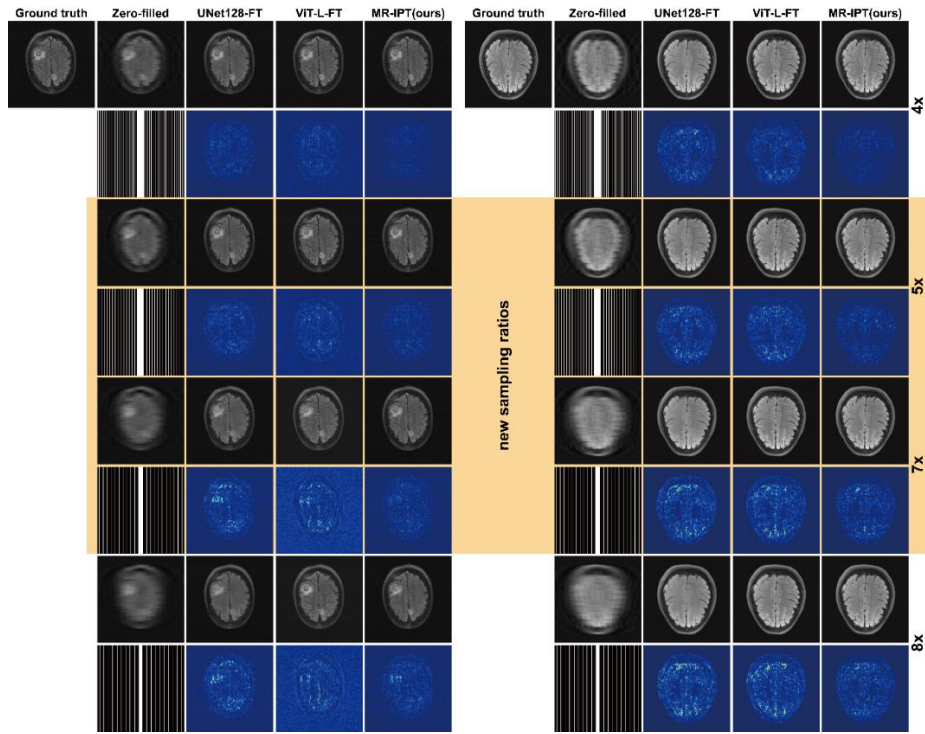

**Figure 4.** Reconstruction comparison across multiple sampling ratios, including unseen configurations. This figure showcases the performance of different models in reconstructing MRI images at various acceleration ratios. Reconstructions highlighted in the yellow block represent new sampling ratios (e.g., 5× and 7×) that were not encountered during pre-training, demonstrating each model's generalization ability. The results highlight MR-IPT's strong adaptability to previously unseen acceleration factors, effectively preserving fine anatomical structures while minimizing artifacts, compared to other baseline methods.

### 2.4 Performance on New Sampling Masks

In this section, we evaluate MR-IPT's adaptability to novel sampling masks. As depicted in Figure 1c, our pretraining strategy includes a diverse set of 1D (Cartesian random, Cartesian equispaced, and 1D Gaussian) and 2D (2D Gaussian) sampling masks. To assess the impact of excluding 2D masks during pretraining, we introduce three MR-IPT-1D variants, trained exclusively on 1D masks. For downstream evaluations, we test all models on accelerated MRI reconstructions using 4× and 8× 2D Gaussian sampling masks.

**Table 3** presents quantitative results comparing standard MR-IPT, MR-IPT-1D, and baseline models UNet128-FT and ViT-L-FT. Notably, both UNet128-FT and ViT-L-FT are pretrained with 2D Gaussian masks, aligning with the standard MR-IPT setup. Surprisingly, despite being trained solely on 1D masks, MR-IPT-1D demonstrates strong generalization and competitive performance on 2D-masked reconstructions. For instance, MR-IPT-level-1D achieves a PSNR/SSIM of 37.99/0.9685 (8×), closely matching MR-IPT-level at 38.94/0.9727 and significantly surpassing UNet128-FT (32.30/0.9310) and ViT-L-FT (33.21/0.9373). Interestingly, when evaluating 1D-masked reconstructions, MR-IPT-1D slightly outperforms the standard MR-IPT due to its specialized pretraining. For example, in 8× Cartesian equispaced reconstruction, MR-IPT-level-1D achieves a PSNR/SSIM of 35.60/0.9562, compared to MR-IPT-level at 35.53/0.9557. **Figure 5** provides qualitative comparisons of reconstructed images, demonstrating that both MR-IPT and MR-IPT-1D generate cleaner error maps for 2D-masked inputs than ViT-L-FT, further validating the robustness and adaptability of the MR-IPT framework.

**Table 3.** Reconstruction performance comparison on unseen sampling masks. MR-IPT-1D models are pretrained exclusively on 1D masks and evaluated on both 1D and 2D-masked reconstructions to assess cross-mask generalization. Standard MR-IPT, UNet128-FT, and ViT-L-FT serve as baselines. Results highlight MR-IPT-1D's ability to adapt to novel 2D masks while maintaining strong performance on 1D-masked reconstructions.

|  | Brain | | | | Knee | |
|---|---|---|---|---|---|---|
|  | 2D Gaussian | | Cartesian Equispaced | | Cartesian Random | |
|  | ACC=4X | ACC=8X | ACC=4X | ACC=8X | ACC=4X | ACC=8X |
| UNet128-FT | 35.62/0.9571 | 32.30/0.9310 | 37.27/0.9653 | 32.03/0.9248 | 32.55/0.8380 | 29.31/0.7566 |
| ViT-L-FT | 37.47/0.9638 | 33.21/0.9373 | 37.63/0.9564 | 33.78/0.9318 | 32.09/0.8032 | 29.33/0.7375 |
| MR-IPT-*type* | 42.32/0.9839 | 38.65/0.9714 | 42.31/0.9827 | 35.34/0.9543 | 34.47/0.8671 | 31.38/0.7942 |
| MR-IPT-*level* | **42.69/0.9849** | **38.94/0.9727** | **42.48/0.9831** | **35.53/0.9557** | **34.52/0.8681** | **31.45/0.7952** |
| MR-IPT-*split* | 42.20/0.9837 | 38.47/0.9707 | 42.06/0.9827 | 35.34/0.9543 | 34.51/0.8678 | 31.44/0.7948 |
| MR-IPT-*type*-1D | 42.01/0.9831 | 37.70/0.9671 | 42.44/0.9830 | 35.54/0.9559 | 34.50/0.8677 | 31.45/0.7950 |
| MR-IPT-*level*-1D | **42.30/0.9839** | **37.99/0.9685** | **42.57/0.9834** | **35.60/0.9562** | **34.62/0.8701** | **31.60/0.7981** |
| MR-IPT-*split*-1D | 42.09/0.9833 | 37.84/0.9679 | 42.53/0.9832 | 35.53/0.9558 | 34.56/0.8680 | 31.53/0.7964 |

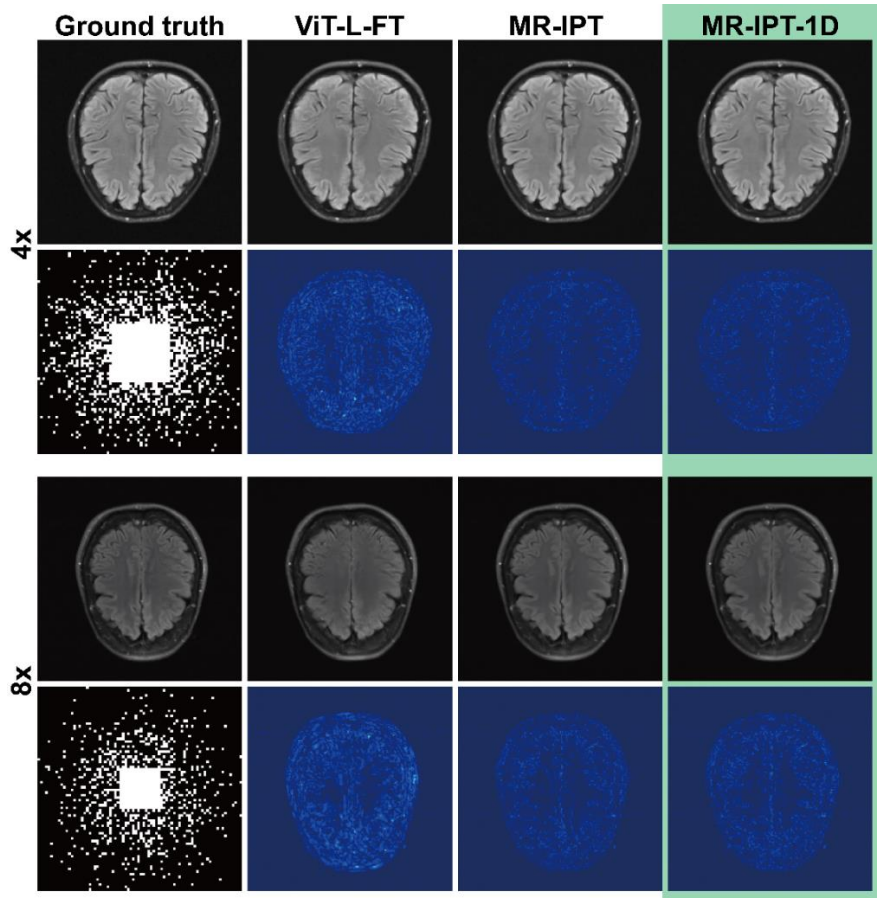

**Figure 5.** Reconstruction comparison using 2D Gaussian sampling masks. Reconstructions highlighted in the green block correspond to MR-IPT-1D models, which were pre-trained exclusively on 1D sampling masks. Despite the absence of 2D masks during pre-training, MR-IPT-1D demonstrates strong generalization capabilities, achieving high-fidelity reconstructions with minimal artifacts, further validating its robustness for new sampling masks.

## 2.5 Zero-shot Performance

In this part, we evaluate the zero-shot performance of MR-IPT in comparison to other models. The quantitative results are summarized in **Table 4**. All models are pretrained on the same dataset before being tested directly on the specified acceleration setups without additional fine-tuning. Across all tested configurations, MR-IPT outperforms both UNet128 and ViT-L by a significant margin, demonstrating superior generalization abaiability. For example, in the 8× Cartesian random sampling test, MR-IPT-level achieves a PSNR/SSIM of 30.65/0.7814, surpassing UNet128 at 27.48/0.7184 and ViT-L at 29.00/0.7296. Similarly, in the 4× Cartesian equispaced test, MR-IPT-level attains a PSNR/SSIM of 39.90/0.9756, outperforming UNet128 at 32.89/0.9315 and ViT-L at 35.73/0.9489. These results highlight the effectiveness of MR-IPT's multi-head-tail design and shared encoder-decoder architecture in adapting to novel sampling patterns without task-specific fine-tuning.

**Table 4.** Zero-shot reconstruction performance comparison. All models are pretrained on the same dataset and evaluated directly on the specified acceleration setups without additional fine-tuning.

|  | Knee | | Brain | | | |
|---|---|---|---|---|---|---|
|  | Cartesian Random | | Cartesian Equispaced | | 2D Gaussian | |
|  | ACC=4X | ACC=8X | ACC=4X | ACC=8X | ACC=4X | ACC=8X |
| UNet128 | 31.08/0.8156 | 27.48/0.7184 | 32.89/0.9315 | 27.93/0.8584 | 33.09/0.9361 | 29.23/0.8888 |
| ViT-L | 31.74/0.7955 | 29.00/0.7296 | 35.73/0.9489 | 30.21/0.9197 | 34.53/0.9399 | 28.55/0.9255 |
| MR-IPT-*type* | 34.07/0.8590 | 30.42/0.7753 | 39.49/0.9740 | 32.08/0.9232 | 39.57/0.9750 | 35.34/0.9528 |
| MR-IPT-*level* | 34.22/0.8635 | 30.65/0.7814 | 39.90/0.9756 | 32.52/0.9288 | 39.56/0.9749 | 35.57/0.9542 |
| MR-IPT-*split* | 34.09/0.8586 | 30.48/0.7739 | 39.46/0.9739 | 31.97/0.9223 | 39.03/0.9729 | 34.84/0.9488 |

## 2.6 Model Stability Regarding Downstream Dataset size

In many real-world scenarios, acquiring large-scale medical imaging datasets for comprehensive training of CNN-based models is challenging. When only a limited number of images are available, fine-tuning MR-IPT on a small dataset becomes a more practical approach. To evaluate MR-IPT's performance under varying dataset sizes, we fine-tuned our model on the fastMRI brain dataset using an 8× Cartesian equispaced mask, with dataset sizes ranging from 10 to 2500. For each size, we randomly sampled subsets from the dataset and repeated the process ten times to assess model performance and stability.

As illustrated in **Figure 6**, both the average performance and stability of MR-IPT improve as dataset size increases. For instance, with just 10 training samples, MR-IPT achieves a PSNR/SSIM of 32.78/0.9316. As the dataset size increases to 2500, the performance improves to 34.05/0.9441, approaching the performance of MR-IPT fine-tuned on the full dataset (35.53/0.9557). These results highlight MR-IPT's ability to leverage large-scale pretraining, maintaining stable and competitive reconstruction quality even in data-constrained settings.

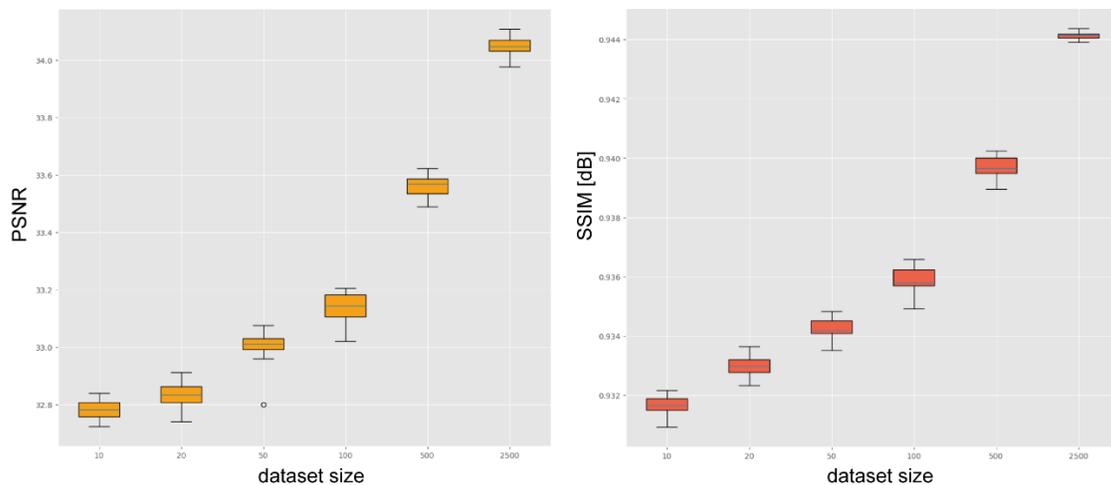

**Figure 6.** MR-IPT performance across different dataset sizes. To assess the impact of dataset size on reconstruction quality, MR-IPT was fine-tuned on the fastMRI brain dataset using an 8× Cartesian equispaced mask, with training subsets ranging from 10 to 2500 images. Each subset was randomly sampled, and the process was repeated ten times to evaluate both model performance and stability. The results indicate that as the dataset size increases, MR-IPT exhibits significant improvements in both reconstruction accuracy and consistency. Even with a limited number of training samples, the model maintains competitive performance, demonstrating its ability to leverage large-scale pretraining effectively and achieve stable, high-quality reconstructions in data-constrained scenarios.

**2.7 Performance Compared to IPT**

Finally, we conducted a comparison between our MR-IPT model and the original IPT framework. Quantitative results are summarized in **Table 5**, with visual reconstruction comparisons shown in **Figure 7**. In IPT, the head-tail setup functions similarly to MR-IPT's split mode, where each unique combination of sampling mask and acceleration ratio corresponds to a dedicated head-tail pair. We evaluate both zero-shot and fine-tuned performance for 4× and 8× undersampling with Cartesian random and Cartesian equispaced masks. Pre-trained on the same dataset, which includes both 1D and 2D sampling masks, MR-IPT generally outperforms IPT across various configurations. For example, in the 4× Cartesian random reconstruction on the knee dataset, MR-IPT achieves a PSNR/SSIM of 34.22/0.8635 (zero-shot) and 34.52/0.8681 (fine-tuned), whereas IPT performs at 31.61/0.8038 (zero-shot) and 32.70/0.8094 (fine-tuned). Similarly, for the 4× Cartesian equispaced reconstruction on the brain dataset, MR-IPT reaches a PSNR/SSIM of 39.90/0.9756 (zero-shot) and 42.48/0.9831 (fine-tuned), compared to IPT's 37.63/0.9580 (zero-shot) and 39.43/0.9640 (fine-tuned).

**Figure 8** illustrates model performance relative to model size for multiple downstream tasks, with the size of each model represented by the size of the corresponding marker. This figure highlights that MR-IPT consistently outperforms IPT, ViT-L-FT, and UNet128-FT, even with similar model sizes. This demonstrates the effectiveness of MR-IPT's lightweight decoder and prompt encoder design, which allows for a larger encoder and thus more effective latent space learning. Overall, our results show that MR-IPT provides robust and stable performance across a wide range of downstream tasks. Notably, it excels in handling new sampling ratios and masks and demonstrates impressive zero-shot generalization as well as stability when trained on limited dataset sizes.

**Table 5.** Comparison of MR-IPT and IPT performance. MR-IPT-level is used as the baseline for evaluation, with both zero-shot and fine-tuned results for various undersampling scenarios across different sampling masks.

|  | Knee - Cartesian Random | | | | Brain – Cartesian Equispaced | | | |
|---|---|---|---|---|---|---|---|---|
|  | ACC=4X | | ACC=8X | | ACC=4X | | ACC=8X | |
|  | PSNR | SSIM | PSNR | SSIM | PSNR | SSIM | PSNR | SSIM |
| Zero-shot | | | | | | | | |
| MR-IPT | 34.22 | 0.8635 | 30.65 | 0.7814 | 39.90 | 0.9756 | 32.52 | 0.9288 |
| IPT | 31.61 | 0.8038 | 29.60 | 0.7382 | 37.63 | 0.9580 | 32.45 | 0.9213 |
| Finetuned | | | | | | | | |
| MR-IPT | 34.52 | 0.8681 | 31.45 | 0.7952 | 42.48 | 0.9831 | 35.53 | 0.9557 |
| IPT | 32.70 | 0.8094 | 30.03 | 0.7469 | 39.43 | 0.9640 | 34.85 | 0.9419 |

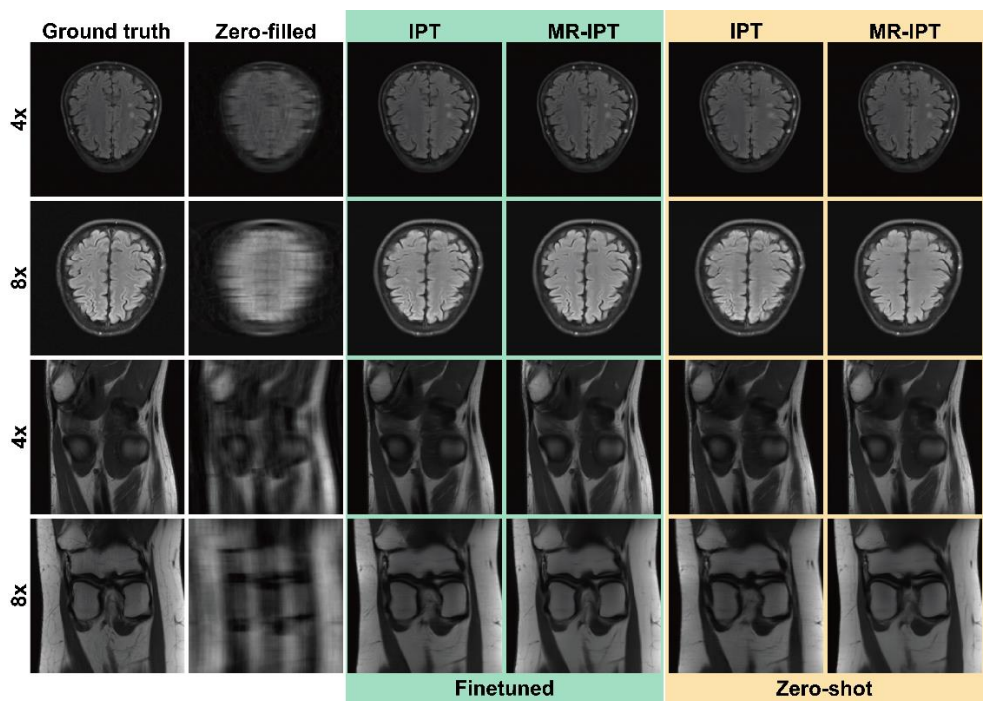

**Figure 7.** Comparison of MR-IPT and IPT reconstructions across different undersampling settings. This figure evaluates the reconstruction performance of MR-IPT against the original Image Processing Transformer (IPT) under both fine-tuned and zero-shot scenarios. Reconstructions in the green block represent fine-tuned comparisons, where both IPT and MR-IPT were trained on 4× and 8× undersampling with Cartesian random and Cartesian equispaced masks. Reconstructions in the yellow block illustrate zero-shot comparisons, where models were tested without additional fine-tuning. The results demonstrate that MR-IPT consistently outperforms IPT, achieving higher reconstruction

fidelity, better structural preservation, and reduced artifacts, highlighting its superior adaptability and generalization capabilities.

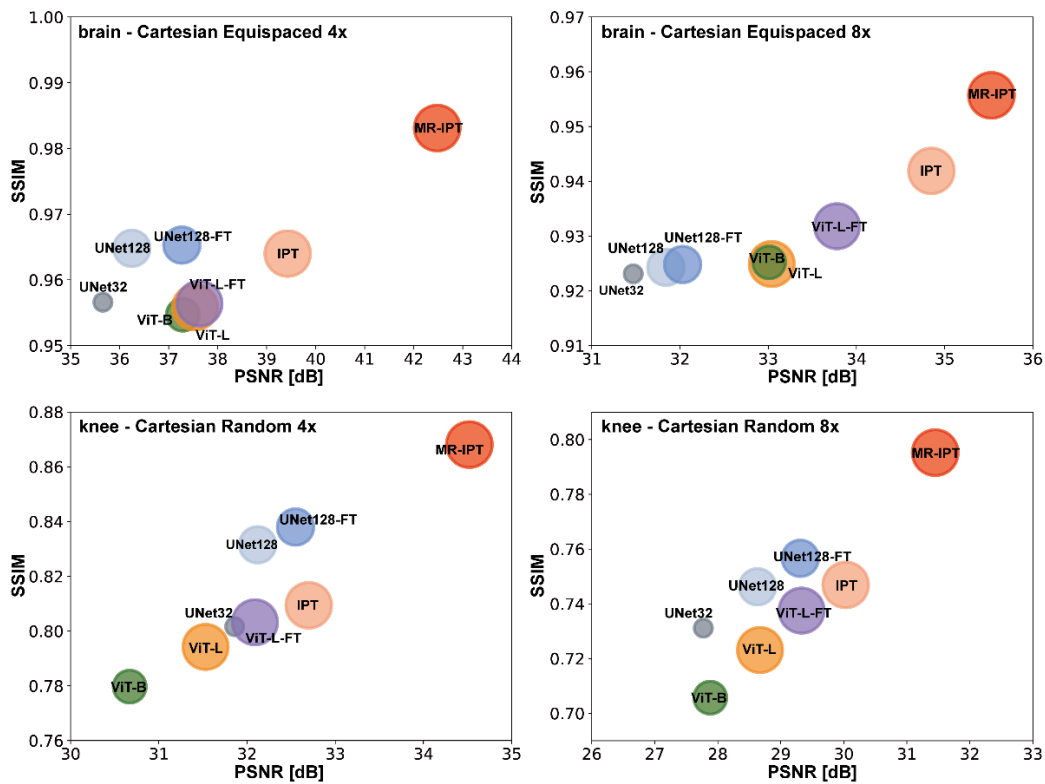

**Figure 8.** Model performance versus size across different reconstruction tasks. Each marker represents a model, with its size proportional to the model's parameter count, illustrating the trade-off between computational complexity and reconstruction performance. The results show that MR-IPT consistently outperforms IPT, ViT-L-FT, and others, even when operating at comparable model sizes. This superior performance is attributed to MR-IPT's lightweight decoder and prompt encoder design, which enables the use of a deeper and more expressive encoder for improved latent space learning without significantly increasing computational cost. The findings highlight MR-IPT's efficiency, scalability, and effectiveness in MRI reconstruction across diverse undersampling conditions.

## 3. Conclusion

In this work, we introduce MR-IPT, a transformer-based framework designed for general accelerated MRI reconstruction. Unlike previous ViT-based models that primarily focus on task-specific reconstruction, MR-IPT is built to fully leverage the potential of ViT through large-scale pre-training. This approach enables MR-IPT to deliver superior performance across diverse accelerated MRI reconstruction tasks. Its multi-head-tail design provides flexibility, supporting a wide range of undersampling masks and acceleration ratios during image degradation. The shared ViT backbone facilitates a unified structure for latent space representation and learning under various sampling conditions. Additionally, the lightweight decoder, deeper encoder, and label prompt encoder contribute to enhanced reconstruction performance compared to conventional ViT models.

MR-IPT consistently demonstrates strong performance compared to traditional MRI reconstruction networks such as UNet and ViT. As illustrated in Table 1 and Figure 2, MR-IPT outperforms these models significantly. For example, in the 4-fold Cartesian equispaced sampling test on the fastMRI brain dataset, MR-IPT-level achieves a PSNR/SSIM of 42.48/0.9831, compared to UNet128 at 36.25/0.9648 and ViT-L at 37.54/0.9558. Even when employing the same pre-training strategies, MR-IPT maintains its advantage, outperforming both UNet128-FT (37.27/0.9653) and ViT-L-FT (37.63/0.9564).

MR-IPT also exhibits stable and superior performance when evaluated on new sampling masks and acceleration ratios, as shown in Tables 2 & 3 and Figures 4 & 5. In the Cartesian equispaced reconstruction test with a new 5× sampling ratio, MR-IPT-level achieves a PSNR/SSIM of 39.92/0.9763, outperforming UNet128-FT (35.52/0.9554) and ViT-L-FT (36.29/0.9490). Remarkably, in the 4× 2D Gaussian sampling reconstruction test, where the model was pre-trained solely with 1D sampling masks, MR-IPT-level-1D achieves a PSNR/SSIM of 42.30/0.9839. This performance is comparable to MR-IPT pre-trained with both 1D and 2D masks (MR-IPT-level at 42.69/0.9849) and superior to UNet128-FT (35.62/0.9571) and ViT-L-FT (37.47/0.9638), both of which included 2D sampling masks during pre-training. These results underscore MR-IPT's adaptability and generalizability in accelerated MRI reconstruction tasks.

Furthermore, MR-IPT demonstrates robust performance in zero-shot learning scenarios, as summarized in Table 4. For instance, in the 8× Cartesian equispaced test, MR-IPT-level achieves a PSNR/SSIM of 32.52/0.9288, outperforming UNet128 (27.93/0.8584) and ViT-L (30.21/0.9197). In situations with limited training data, MR-IPT maintains stable and high-quality reconstruction performance, as shown in Figure 6. These results highlight MR-IPT's ability to leverage large-scale pre-training effectively, ensuring competitive reconstruction quality even in data-constrained settings.

When compared to the original IPT framework, MR-IPT consistently exhibits superior performance in both zero-shot and fine-tuned scenarios, as demonstrated in Table 5 and Figure 7. For example, in the 4× Cartesian equispaced reconstruction of the brain dataset, MR-IPT achieves a PSNR/SSIM of 39.90/0.9756 (zero-shot) and 42.48/0.9831 (fine-tuned), compared to IPT's 37.63/0.9580 (zero-shot) and 39.43/0.9640 (fine-tuned). Comparative studies (Figure 8) further reveal that MR-IPT outperforms IPT, ViT-L-FT, and UNet128-FT even with similar model sizes. This superior performance is attributed to MR-IPT's lightweight decoder and prompt encoder design, which enables a larger encoder for more effective latent space learning.

Despite its strong performance, this study has several limitations. First, the dataset used for pre-training is limited to accelerated MRI data. Although previous studies have shown that pre-training on general image datasets can enhance accelerated MRI reconstruction,[30] our pre-training dataset remains confined to medical imaging. Future work should investigate the benefits of incorporating more diverse datasets and explore the potential of merging multiple datasets to scale up pre-training, potentially improving model performance. Second, our downstream tasks are primarily focused on image reconstruction. Recent works like MAE[27] and SAM[28] have demonstrated the versatility of pre-trained encoders for tasks such as classification and object detection. Further research is needed to evaluate MR-IPT's performance in other downstream applications, including segmentation and disease detection. Additionally, integrating advanced techniques such as denoising diffusion models[38] and flow matching[39, 40] holds potential for developing more generalized multimodal frameworks in medical imaging. Integrating MR-IPT with these methods could broaden its applicability beyond MRI reconstruction. Finally, MR-IPT currently employs a simple L1 loss function during training. Future studies will explore more sophisticated objective functions, such as SSIM loss and contrastive loss, to further enhance reconstruction quality.

In conclusion, we present MR-IPT, a ViT-based framework for general accelerated MRI reconstruction. MR-IPT demonstrates superior performance across various sampling setups, including new sampling masks and ratios, zero-shot learning, and limited dataset scenarios. By leveraging a multi-head-tail structure and a shared backbone design, MR-IPT exhibits strong adaptability to diverse acceleration configurations. Its efficient latent representation learning and large-scale pre-training capabilities highlight the potential of ViT-based architecture in advancing medical imaging deep learning models. This approach represents a significant step forward in the development of generalizable, high-performance deep learning models for medical imaging, with the potential to advance both clinical applications and future AI-driven diagnostic tools.

## 4. Experimental Section
### 4.1 MR-IPT Model Structure and Details

The MR-IPT framework builds upon a modified IPT structure, specifically designed to accommodate a wide range of accelerated MRI reconstruction tasks. The architecture of MR-IPT is illustrated in Figure 1, showcasing its five core components: heads, tails, a prompt encoder, a shared encoder, and a lightweight decoder.

Each head consists of a 3×3 convolutional layer followed by two 5×5 residual blocks, enabling robust feature extraction from accelerated images. The tail, responsible for image reconstruction, includes a 3×3 upsampling convolutional block followed by another 3×3 convolutional layer to refine output quality. The prompt encoder, inspired by the design of the Segment Anything Model (SAM),[28] utilizes sparse embeddings to project acceleration information into the same dimensional space as image embeddings, enhancing task-specific feature representation.

For the shared encoder, we adopted a 24-layer ViT equipped with W-MSA,[34] which effectively captures global contextual information across multiple layers. This design mirrors the heavy-weight encoder/light-weight decoder paradigm seen in MAE,[27] balancing computational efficiency with strong representational capacity. The 2-layer lightweight decoder incorporates two-way cross-attention mechanisms between image patch tokens and prompt tokens, facilitating effective feature refinement and image reconstruction.

During pre-training, MR-IPT is exposed to diverse acceleration ratios (2×, 4×, 6×, 8×, and 10×) and a variety of sampling masks. The 1D sampling masks include Cartesian random, Cartesian equispaced, and 1D Gaussian, while the 2D sampling mask is based on 2D Gaussian distributions. We developed three MR-IPT variants to explore different aggregation strategies: (1) MR-IPT-type: Heads and tails are grouped based on acceleration ratios, with each head-tail pair specialized for different sampling masks; (2) MR-IPT-level: Heads and tails are aggregated based on sampling masks, allowing each pair to handle various acceleration ratios; (3) MR-IPT-split: A dedicated head-tail pair is assigned to each unique combination of sampling mask and acceleration ratio.

For downstream tasks, the appropriate head-tail pair is selected based on the specific sampling configuration. In cases involving unseen acceleration ratios (e.g., 5×), we utilize the head-tail pair trained for the nearest higher ratio (e.g., 6×). For MR-IPT-1D models, pre-training is limited to 1D masks, and during downstream evaluations, head-tails trained with Cartesian random masks are preferred due to their robustness against diverse sampling strategies.

MR-IPT was implemented using PyTorch[41] and trained on systems equipped with either an NVIDIA RTX 3090 Ti (24GB VRAM) or an NVIDIA RTX 4090 (24GB VRAM). The model optimization follows an Adam optimizer[42] with a learning rate of 1e-5. The training protocol includes a pre-training for 5 epochs on the large-scale dataset, then a fine-tuning for 15 epochs for each downstream reconstruction task.
The training objective for MR-IPT is L1 loss, defined as:

$$\mathcal{L} = \sum_{i=1}^{N} \left\| MRIPT(x_{accelerated}^i) - x_{clean} \right\|_1 \tag{1}$$

where $x_{accelerated}^i$ denotes the accelerated image for sampling task $i$, and $x_{clean}$ denotes the fully sampled clean ground truth image.

## 4.2 Datasets and Image Processing

For large-scale pre-training, we utilized the RadImageNet dataset,[35] which supports the findings of this study and is publicly available at https://www.radimagenet.com/. Specifically, we employed the MRI subset, comprising 672,675 images, which were split into a 9:1 ratio, resulting in 605,408 images for training and 67,267 for validation. For downstream task evaluations, we used the fastMRI dataset,[36] which is openly accessible at https://fastmri.med.nyu.edu/. For fastMRI knee dataset, we used the training set including 34742 images for training. For testing, we used the validation set including 7135 images. For fastMRI brain dataset, we used the training set including 70748 images for training. For testing, we used the test set including 8852 slices. To standardize the data, all images were resized to 224×224 pixels with pixel intensity values normalized within the range of [0, 1]. Given the variability in image dimensions across the fastMRI datasets, we first applied center cropping to reduce the images to 320×320 pixels, ensuring the preservation of central anatomical features, and subsequently resized them to the target dimensions for model training and evaluation.

## 4.3 Evaluation Metrics

To comprehensively assess reconstruction performance, we adopted two widely used quantitative metrics during our reconstruction performance evaluation comparisons: peak signal-to-noise ratio (PSNR) and structural similarity index (SSIM):

$$PSNR(x, x_{clean}) = 10 \log_{10} \frac{\max(x_{clean})^2}{MSE(x, x_{clean})} \tag{2}$$

where $MSE(x, x_{clean})$ is the mean squared error between the reconstructed image $x$ and the fully sampled ground truth clean image $x_{clean}$.

$$SSIM(x, x_{clean}) = \frac{(2\mu_x \mu_{x_{clean}} + C_1)(2\sigma_{xx_{clean}} + C_2)}{(\mu_x^2 + \mu_{x_{clean}}^2 + C_1)(\sigma_x^2 + \sigma_{x_{clean}}^2 + C_1)} \tag{3}$$

where $\mu_x$, $\mu_{x_{clean}}$, $\sigma_x^2$, and $\sigma_{x_{clean}}^2$ are the mean and variance of reconstructed image $x$ and fully-sampled clean image $x_{clean}$, respectively. $\sigma_{xx_{clean}}$ is the covariance of $x$ and $x_{clean}$. $C_1 = (0.01L)^2$, $C_2 = (0.03L)^2$, where $L$ is the dynamic range of the pixel-values.

**Acknowledgements**

This work is supported by the Rajen Kilachand Fund for Integrated Life Science and Engineering. We would like to thank the Boston University Photonics Center for technical support.

**Equal Contribution**

Guoyao Shen and Mengyu Li contributed equally to this work.

**Conflict of Interests**

The authors have no conflicts of interest to disclose.

**Data Availability**

The RadImageNet dataset that support the findings of this study are openly available at: https://www.radimagenet.com/. The fastMRI dataset that support the findings of this study are openly available at: https://fastmri.med.nyu.edu/. Our code is available at: https://github.com/GuoyaoShen/MR-IPT.

**Author Contributions**

**Guoyao Shen:** Conceptualization (lead); Methodology (equal); Software (equal); Formal Analysis (equal); Writing – Original Draft (lead). **Mengyu Li:** Methodology (equal); Software (equal); Formal Analysis (equal); Writing – Original Draft (supporting). **Stephan Anderson:** Methodology (supporting); Writing – Review & Editing (supporting). **Chad W. Farris:** Conceptualization (supporting); Formal Analysis (supporting); Writing – Review & Editing (supporting). **Xin Zhang:** Conceptualization (lead); Methodology

(equal); Software (equal); Formal Analysis (equal); Writing – Review & Editing (lead); Project Administration (lead); Funding Acquisition (lead).